\documentclass[pra,preprint,tightenlines,showpacs,twocolumn,10pt]{revtex4} 
\usepackage{epsfig} 

\begin{document}  

\title{ Relativistic transfer ionization and the Breit interaction }


\author{O.\ Yu.\ Andreev$^1$, E.\ A.\ Mistonova$^1$ and A. B. Voitkiv$^2$}
\affiliation{$^1$ Physics Department,
             St.\ Petersburg State University, St.\ Petersburg, Russia \\ 
             $^2$ Max-Planck-Institut f\"ur Kernphysik, 
             Heidelberg, Germany }


\begin{abstract}

We consider correlated transfer ionization in relativistic collisions between a highly charged ion and a light atom. 
In this process two quasi-free electrons of the atom 
interact with each other during the short collision time 
that results in capture of one of them by the ion and emission of the other. We show that this process is strongly influenced 
by the generalized Breit interaction already 
at modest relativistic impact energies.  

\end{abstract}


\maketitle

Charged particles moving with velocities much smaller than 
the speed of light $c$ ($c \approx 137$ a.u.) 
interact with each other mainly via the Coulomb force. When the velocities increase and approach the speed of light the electromagnetic fields generated by the particles start to noticeably deviate from the Coulomb law which in turn influences the form of their interaction. Nevertheless, in many cases the effect of the relativistic corrections to the Coulomb force remains quite modest unless 
the impact energies reach extremely high values. 

For instance, the total cross section for ionization of a light atom 
by collision with an ion is proportional 
to $ \ln a v + \ln \gamma - v^2/2c^2 $ (see e.g. \cite{fano}), 
where $v$ is the collision velocity (in a.u.), 
$ \gamma = 1/\sqrt{1-v^2/c^2} $ and $a \sim 1$ -- $10$ is a quantity whose value depends on the properties of the atom. The last two terms  
arise due to the relativistic corrections to the Coulomb interaction between the ion and the atomic electron. In order that these terms 
change the cross section by $2$ or $5$ times one should have 
$\gamma \sim 10^2$ or $10^8$, respectively,  
that corresponds to impact energies of the order of  
$100$ GeV/u or $10^5$ TeV/u.     

When the interacting particles are electrons   
the (leading) relativistic correction to the Coulomb force  
is given by the generalized Breit interaction (GBI) \cite{gbi}. 
The latter is usually derived within the scope of Quantum Electrodynamics 
appearing as a result of the exchange of a transverse photon 
between the electrons. However, as a rule, 
the contribution of the GBI remains a small correction 
to that due to the Coulomb force. 

For ionization of a light atom by high-energy electrons  
the situation with importance of the relativistic correction  
is quite similar to that in ionization by the ion impact. 
In particular, in order that the GBI increases 
the total cross section for ionization by 
$2$ or $5$ times the electron impact energy 
should reach enormous values 
($\sim 100$ MeV or $10^2$ TeV, respectively). 

Compared to ionization of light atoms the GBI 
is much more important for excitation and 
'ionization' of very highly charged ions 
(like e.g. U$^{91+}$(1s)) by electron impact 
\cite{fontes}, \cite{gumberidze}. However, even 
in such a case, in order that the GBI could 
strongly dominate these processes 
the impact energy also has to be extremely high.   

The electron-electron interaction plays   
an important role in the projectile-electron loss 
\cite{abv-buch}. In this process a partially 
stripped projectile-ion looses 
an electron due to the collision with a neutral atom. 
The interaction acting between the electrons 
of the ion and atom may result in the simultaneous 
electron loss from the ion and ionization of the atom. 
The effect of the GBI in such a process, however, 
remains modest no matter how 
high is the impact energy \cite{mut-ioniz}.    

The GBI can influence dielectronic recombination.   
However, it can dominate this process 
only provided some special transitions are considered 
\cite{dr-2} and remains on overall 
more or less a minor correction to the Coulomb force.  

Here we consider correlated transfer-ionization 
in relativistic collisions between a highly charged ion (HCI) 
and a light atom. In this process two atomic electrons 
interact with each other during 
the very short collision time that results in  
capture of one of them by the ion 
and emission of the other. 
It turns out that, in contrast to the processes mentioned above, 
the correlated transfer ionization is profoundly influenced 
by the GBI already at very modest 
relativistic impact energies. 

Note that while transfer ionization in the relativistic domain 
was not yet explored, the different aspects (correlated and uncorrelated) of this process in nonrelativistic collisions ($v \ll c$) with low charged ions were extensively studied 
(see \cite{mergel}-\cite{gulyas} and references therein).   

Atomic units ($\hbar = m_e = e =1 $) are used throughout  
except where otherwise stated.    

Let us consider the collision between a highly charged bare ion with a charge $Z_i$ and a light atom with atomic number $Z_a$ ($Z_a \ll Z_i$ ) which has two K-shell electrons. Our consideration employs the semi-classical approximation in which the electrons are treated as quantum particles whereas the heavy nuclei are described classically.
It is convenient to perform the basic consideration using the rest frame of the HCI and take its position as the origin. In this frame the nucleus of the atom moves along a straight-line classical trajectory,  
${\bf R}(t)={\bf b}+{\bf v} t$ , where ${\bf b}$, ${\bf v}$ and $t$ 
are the impact parameter, the velocity of the atom and the time, respectively. Note that in relativistic collisions 
$v$ is much larger than the typical orbiting velocities 
$\sim Z_a$ of the electrons in the atom. 

If two free electrons are incident on an ion  
one of them can be captured into a bound state 
of the ion due to the interaction with 
the other electron which carries away 
the energy excess in such a three-body recombination process. 
In the ion-atom collision the electrons are initially 
bound in the incident atom. However, if the change 
in their momenta in the collision 
is much larger than their orbiting momenta in the atom
they can be regarded as (quasi) free. 
Such an approximation is very well established 
in the calculations of radiative and nonradiative capture 
of a single electron (see e.g. \cite{imp-like}) 
and is also valid for other collision processes characterized 
by large change in the momentum of the atomic electrons 
(see e.g. \cite{2el-1ph} - \cite{we-excitation}).     
In our case the change in the momenta 
of the electrons is $ \sim \gamma v $ ($\gamma = 1/ \sqrt{1-v^2/c^2}$), 
it greatly exceeds their orbiting momenta $ \sim Z_a $ in the atom 
and the approximation of quasi-free electrons is well justified.  

The correlated transfer ionization proceeds via the (single) interaction 
between the quasi free atomic electrons during the short collision time  
when the ion and atom are very close to each other.   
The transition amplitude for this process, 
which is of the first order in 
the electron-electron interaction,   
consists of the 'direct' and 'exchange' contributions. 
The 'direct' one is given by \cite{we-excitation} 
\begin{eqnarray}
a_{fi}^d({\bf b})=-\frac{i}{c^2} \sum_{\mu=0}^3 \int d^4 x j_{1 \mu} (x, {\bf b}) 
A^{\mu} (x,{\bf b}).
\label{e1}
\end{eqnarray} 
Here, $j_{1\mu}(x)$ ($\mu=0,1,2,3$) is electromagnetic transition four-current generated by one of the electrons in the collision at a space-time point $x=(x_0, {\bf x})$ and $A^{\mu} (x,{\bf b})$ is the four-potential of the electromagnetic field created by the other electron at the same point. It satisfies the Maxwell equations   
\begin{eqnarray}
(\frac{ \partial^2 }{ \partial x^2_0 }-\Delta) A^{\mu} (x,{\bf b}) = 
\frac{4 \pi}{c} j_2^{\mu} (x,{\bf b}), 
\label{e2}
\end{eqnarray} 
where $j_2^{\mu} (x,{\bf b})$ is the transition 
four-current generated by the other electron in the collision. 

The field of the HCI, because of its strength, has crucial impact 
on the motion of the electrons not only in the final but also in the initial channel \cite{imp-like}, \cite{we-excitation}, \cite{we-hci}.  
Taking into account the effect of the HCI's field on the initial state in 
the impulse approximation and using Eqs.(\ref{e1})-(\ref{e2}) 
one can show that the Fourier transform of (\ref{e1}) into the momentum space is given by      
\begin{eqnarray} 
S_{fi}^d({\bf q}_{\perp}) &=& -\frac{i}{2 \pi \gamma v } \int d^3 
\mbox{\boldmath$\kappa$} \, 
\xi_a(\mbox{\boldmath$\kappa$}, {\bf q}-\mbox{\boldmath$\kappa$}) 
\nonumber \\
& & \times \int d^3 {\bf r}_1 \int d^3 {\bf r}_2 
\left( L_{\rm Coul} + L_{\rm GBI} \right),   
\label{e3}
\end{eqnarray} 
where $\xi_{a}$ is the momentum distributions of the electrons 
in the initial atomic state as viewed in the rest frame 
of the atom (the atomic Compton profile) and the terms  
\begin{eqnarray}
L_{\rm Coul} & = & 
\chi_b^{\dagger} ({\bf r}_2) \chi^{(+)}_{{\bf p}_2 }({\bf r}_2)
\frac{ 1 }{ r_{12} }
\chi^{(-)\dagger}_{ {\bf p} } ({\bf r}_1) 
\chi^{(+)}_{{\bf p}_1 }({\bf r}_1) 
\nonumber \\  
L_{\rm GBI} & = &  -\chi^{\dagger}_{b} ({\bf r}_2) 
\mbox{\boldmath$\alpha$}_2 \chi^{(+)}_{{\bf p}_2}({\bf r}_2)
\frac{e^{i K_0 r_{12}}}{r_{12}}
\chi^{(-) \dagger}_{{\bf p} } ({\bf r}_1) \mbox{\boldmath$\alpha$}_1 
\chi^{(+)}_{{\bf p}_1 }({\bf r}_1)
\nonumber \\
&+&\chi^{\dagger}_{b} ({\bf r}_2) \chi^{(+)}_{{\bf p}_2 }({\bf r}_2)
\frac{e^{i K_0 r_{12}}-1}{r_{12}}
\chi^{(-) \dagger }_{{\bf p} } ({\bf r}_1)  
\chi^{(+)}_{{\bf p}_1 }({\bf r}_1)   
\label{e4}
\end{eqnarray}
arise due to the Coulomb and GBI parts, respectively, 
of the electron-electron interaction. 
In (\ref{e4}) $\mbox{\boldmath$\alpha$}_j$ are the Dirac matrices for the 
$j$-th electron ($j=1,2$), 
$\chi^{(+)}_{{\bf p}_1 }$ and $\chi^{(+)}_{{\bf p}_2 }$ 
with ${\bf p}_1=\mbox{\boldmath$\kappa$}_{\perp} + 
\gamma ( \kappa_z + v \varepsilon^{(1)}_a/c^{2}){\bf v}/v $   
and ${\bf p}_2={\bf q}_{\perp} - \mbox{\boldmath$\kappa$}_{\perp} + 
\gamma (q_z  - \kappa_z + v \varepsilon^{(2)}_a/c^{2}){\bf v}/v $ 
describe the motion of the incident electrons in the field 
of the HCI ($\varepsilon^{(1)}_a$ and $\varepsilon^{(2)}_a$ are 
the electron energies in the initial atomic state), 
and $\chi_b$ and $\chi^{(-)}_{{\bf p} }$ are the final 
states of the captured and emitted electrons, respectively. 
${\bf q} = ({\bf q}_{\perp}, q_z)$
is the momentum transfer in the collision, 
where ${\bf q}_{\perp}$ and $q_z=(\varepsilon_p + \varepsilon_{b}-\gamma \varepsilon_a)/ \gamma v $ are its transverse and longitudinal parts,  respectively, $\varepsilon_b$ is the energy of the captured electron, 
$\varepsilon_p$ is the energy of the emitted electron,   
$ \varepsilon_a =  \varepsilon^{(1)}_a+\varepsilon^{(2)}_a$  and 
$K_0=|\varepsilon_b-\gamma \varepsilon_{a}^{(1)}- \gamma v \kappa_z|/c$.  

Eqs.(\ref{e3})-(\ref{e4}) were derived using 
the Feynman gauge. Note that in a gauge-invariant treatment  
the same result is obtained no matter which gauge 
(e.g. Feynman or Coulomb) is employed \cite{lindgren}. 

Due to the term  
$\xi_a(\mbox{\boldmath$\kappa$}, {\bf q}-\mbox{\boldmath$\kappa$})$ 
the integrand in Eq. (\ref{e3}) 
becomes very small if $|\mbox{\boldmath$\kappa$}|$ and/or  
$ |{\bf q}-\mbox{\boldmath$\kappa$}|$ substantially 
exceed the typical orbiting momenta 
($\sim Z_a$) of the electrons in the atom. 
Therefore, taking into account that $Z_a \ll v$ and neglecting 
the binding energy of the electrons in the atom compared to their rest energy, $\varepsilon^{(1,2)}_a \approx mc^2$, 
one can set $ {\bf p}_1 = {\bf p}_2 = m \gamma {\bf v} $ and 
$K_0=| \varepsilon_b-\gamma mc^2 |/c$.  

The 'exchange' contribution $S_{fi}^e$ to the transition amplitude 
can be obtained from (\ref{e3})-(\ref{e4}) 
by interchanging the electrons in the initial (or final) states. 
The amplitude $ S_{fi} $ for the correlated transfer ionization  
is $ S_{fi} = S_{fi}^d - S_{fi}^e$.  

The cross section for the transfer ionization, 
differential in energy and angle of the emitted electron, reads 
\begin{eqnarray}
\frac{d^{2}\sigma}{d\varepsilon  
\sin \vartheta_p d\vartheta } = 
\int_{0}^{2\pi} d \varphi_p \int d^2 {\bf q}_{\perp} |S_{fi}({\bf q}_{\perp})|^2 
\label{cross-section}
\end{eqnarray} 
where $ \vartheta_p $ is the polar emission angle. 
In (\ref{cross-section})
the integration runs over the transverse part 
of the momentum transfer and the azimuthal angle 
$\varphi_p$ of the emission.   

\begin{figure}[t] 
\vspace{0.25cm}
\begin{center}
\includegraphics[width=0.5\textwidth]{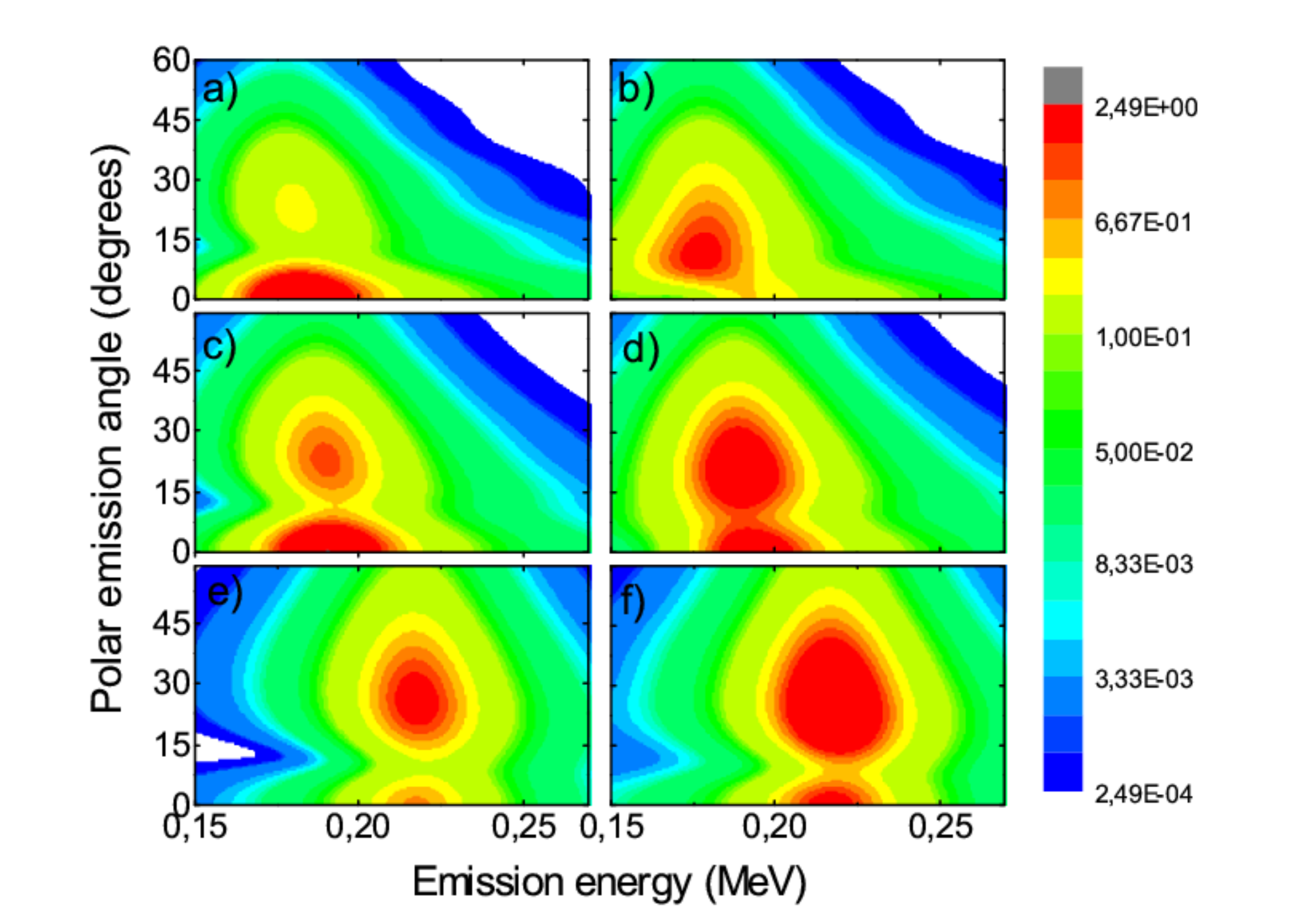}
\end{center}
\vspace{-0.5cm}  
\caption{ \footnotesize{ 
Energy-angular spectrum (in mb/(MeV rad)) 
of the emitted electrons in $152$ MeV/u 
X$^{Z_i+}$+C $\longrightarrow$ X$^{(Z_i-1)+}$(1s)+C$^{2+}$+e$^{-}$ collisions (v=70 a.u.). $Z_i=30$ (a,b), $Z_i=40$ (c, d), $Z_i=60$ (e,f). 
The spectrum is given in the rest frame of the HCI, 
the emission angle is counted from the velocity of the atom. 
The results shown in the right (left) column are obtained using 
the full relativistic (Coulomb) electron-electron interaction. 
}}  
\label{figure1} 
\end{figure}

\begin{figure}[t] 
\vspace{0.25cm}
\begin{center}
\includegraphics[width=0.5\textwidth]{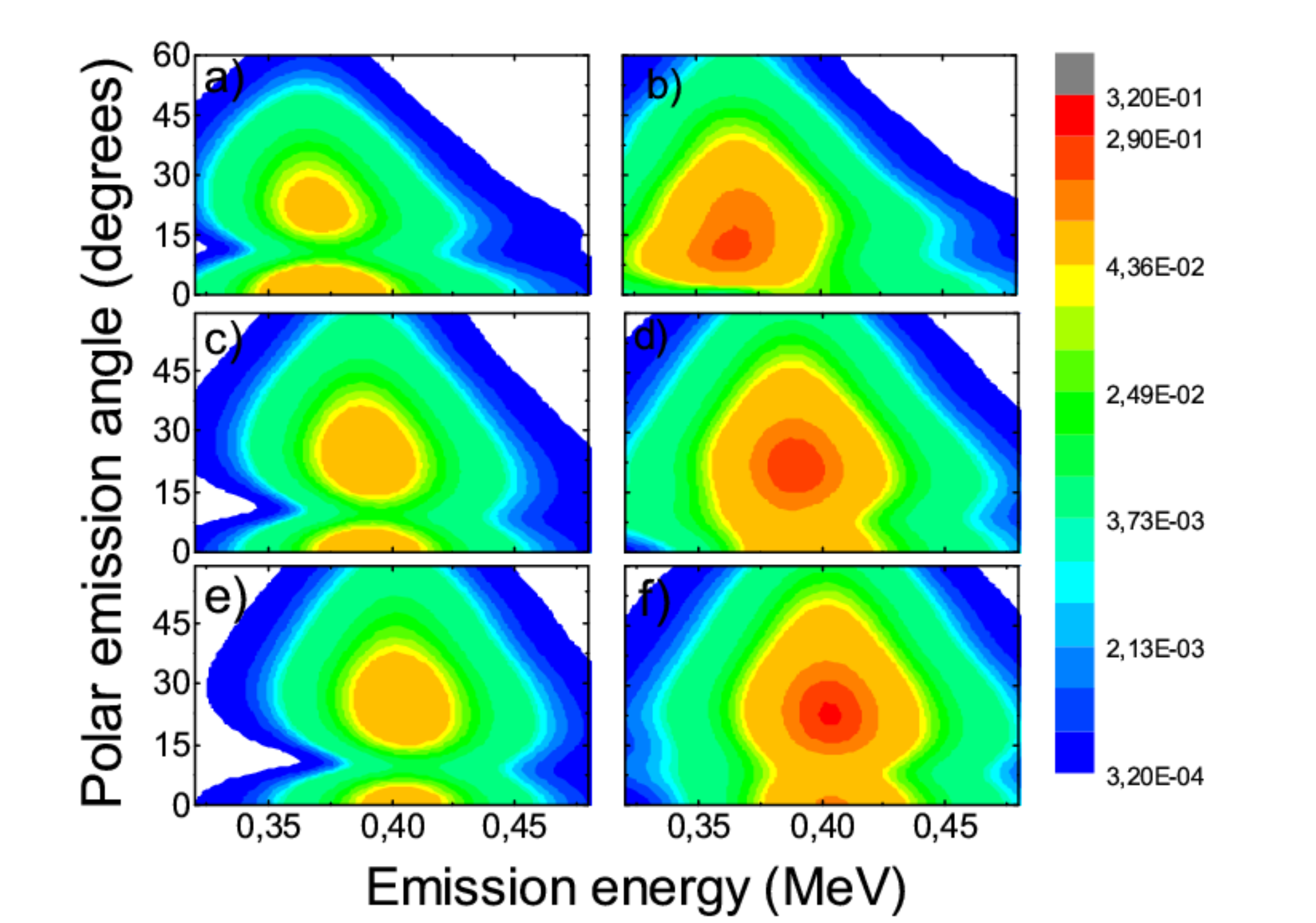}
\end{center}
\vspace{-0.5cm}  
\caption{ \footnotesize{ 
Same as in figure \ref{figure1} but for 
an impact energy of $304$ MeV/u ($v=90$ a.u.) 
and $Z_i=50$ (a,b), $Z_i = 63$ (c, d) and  $Z_i=70$ (e,f). }}  
\label{figure2} 
\end{figure}

Our results for the cross section (\ref{cross-section}) 
(which determines the energy-angular spectrum of the emitted electrons)      
are shown in figures (\ref{figure1})-(\ref{figure2}) 
for collisions of carbon atoms with different HCIs 
at two impact energies.   
The K-shell atomic electrons are mostly 
tightly packed and, as a result, 
have the strongest interaction with each other. 
Therefore, only these electrons were taken into account in our calculations \cite{elect_pairs}. 
Note also that the results shown in the figures were computed by describing the states  $ \chi_b $, $ \chi_{\bf{p}_1} $, $ \chi_{\bf{p}_2} $ and $ \chi_{\bf{p}} $ fully relativistically 
(as a Dirac bound state and Dirac-Coulomb continuum states, respectively).  

The results, displayed in the left column of these figures, 
were obtained by neglecting the GBI. In such an approximation the calculated spectra contain two maxima centered at 
$ \vartheta_p = 0^{\circ} $ and $ \vartheta_p \approx 25^{\circ} $. The shape of these spectra is quite similar to that obtained for the transfer ionization in 
nonrelativistic ($v \ll c$, $Z_i \ll c $) collisions with HCIs \cite{we-hci}-\cite{we-hci-1}. 

In the nonrelativistic case the maximum at $ \vartheta_p = 0^{\circ} $ is caused by the so called electron-electron Auger (EEA) mechanism of transfer ionization \cite{we-hci-1} (which was proposed in \cite{we-eea} and confirmed experimentally in \cite{daniel}). 
The other maximum appears due to the distortion of the motion of both electrons in the continuum by the Coulomb field of the ion \cite{we-hci-1} 
(the corresponding mechanism might be termed 
the electron-electron Coulomb distortion (EECD) mechanism). In 
contrast to the EEA maximum it vanishes if the charge 
of the ion becomes small ($Z_i \ll v$) \cite{eet}.    

The right column in figures (\ref{figure1})-(\ref{figure2}) 
presents the results obtained when 
the full relativistic form of the electron-electron interaction 
is taken into account. Comparing the right and left columns 
one can conclude that the GBI strongly 
influences the shape of the electron emission. 

If the charge of the HCI is noticeably smaller than the collision velocity $v$ the full relativistic calculation predicts only one maximum in the spectrum of the emitted electrons (see the section (b) of 
figures (\ref{figure1})-(\ref{figure2})). Its position differs from 
those of the maxima obtained neglecting the GBI. 
When, at a fixed velocity $v$, the charge of the HCI increases 
this maximum stretches expanding to smaller and larger emission angles 
$\vartheta_p$ (not shown in the figures) and eventually 
splits into two well separated maxima centered at 
$\vartheta_p = 0^{\circ} $ and $\vartheta_p \approx 25^{\circ} $
(see the sections (d) and (f) of the figures). 
The maximum at $\vartheta_p = 0^{\circ} $ is caused by the EEA  
mechanism whereas the other one appears 
due to the EECD. Similarly to the nonrelativistic case 
\cite{we-hci}-\cite{we-hci-1} the relative importance 
of the EEA (EECD) decreases (increases) 
when the HCI's charge grows.  

In order to get insight into the role of the distortion of the electron motion by the Coulomb field of the HCI we have performed two more sets of calculations. In one of them the functions $ \chi_{\bf{p}_1} $, $ \chi_{\bf{p}_2} $ and $ \chi_{\bf{p}} $ were approximated by the relativistic (Dirac) plane waves. In the second the function $ \chi_{\bf{p}_1} $ for the electron to be captured was taken as the Dirac-Coulomb continuum state whereas the functions $ \chi_{\bf{p}_2} $ and $ \chi_{\bf{p}} $ 
for the other electron were again approximated by the plane waves. 

According to such calculations the effect of the relativistic part of 
the electron-electron interaction on the shape of the emission spectra is weak. In particular, in all these calculations the spectrum has only one maximum centered at $\vartheta_p = 0^{\circ} $ no matter whether the electron-electron interaction is taken in its full relativistic form or approximated by the Coulomb interaction.  
It means that the effect of the GBI in the correlated transfer ionization is intimately connected with the distortion of the motion of {\it both} electrons in the Coulomb field of the highly charged ion. 

The GBI not only qualitatively changes the shape 
of the electron emission pattern but also has a strong impact 
on the absolute values of the cross section. In Table~\ref{tab1} we present the total cross section for the correlated transfer ionization. 
This cross section was calculated by neglecting ($\sigma_{coul}$) and including ($\sigma_{full}$) the GBI. 

It follows from the Table that the GBI substantially influences the total cross section already at a modest relativistic impact energy of 152 MeV/u ($v=70$ a.u.) where the corresponding collisional Lorentz factor is still not far from unity ($\gamma = 1.16$ ). 
Moreover, the GBI may increase the cross section 
by more than order of magnitude at an impact energy of  
$630$ MeV/u ($v=110$ a.u.) for which the Lorentz factor 
remains well below $2$ ($\gamma = 1.68$). This could be compared with electron impact 'ionization' of U$^{91+}$(1s) where an order-of-magnitude increase in the total cross section due to the GBI would be reached at impact energies $ \sim 10^{15} $ eV corresponding 
to the Lorentz factor of $ \sim 10^9 $ . 

\begin{table}
\caption{ The total cross section for the correlated transfer ionization. The first and second columns show the collision velocity and the colliding system, respectively, the third and fourth columns display the cross section calculated by ignoring ($\sigma_{coul}$) and including 
($\sigma_{full}$) the GBI, the last column gives the ratio $R=\sigma_{full}/\sigma_{coul}$.}
\label{tab1}
\begin{tabular}{|c|c|c|c|c|c|c}
\hline
$v$ (a.u.) & Colliding particles &  $\sigma_{coul}$ (mb) & $\sigma_{full}$ (mb)  & 
$R$ \\ 
\hline 
70  & Zn${}^{30+}+$C &  2.74 & 4.93  & 1.8 \\
    & Zr${}^{40+}+$C &  5.61 & 9.51  & 1.7 \\
    & Nd${}^{60+}+$C &  8.91 & 17.74 & 2.0 \\
\hline
90  & Sn${}^{50+}+$C &  0.47 & 1.55 & 3.3 \\
    & Eu${}^{63+}+$C &  0.74 & 2.32 & 3.1 \\ 
    & Yb${}^{70+}+$C &  0.81 & 2.67 & 3.3 \\
\hline
110 & Yb${}^{70+}+$C &  0.04 & 0.39 & 8.3 \\
    & At${}^{85+}+$C &  0.06 & 0.49 & 7.7 \\ 
    & U${}^{92+}+$C  &  0.07 & 0.92 & 13.1 \\
\hline
\end{tabular}
\\
\end{table} 

\begin{figure}[t] 
\vspace{0.25cm}
\begin{center}
\includegraphics[width=0.5\textwidth]{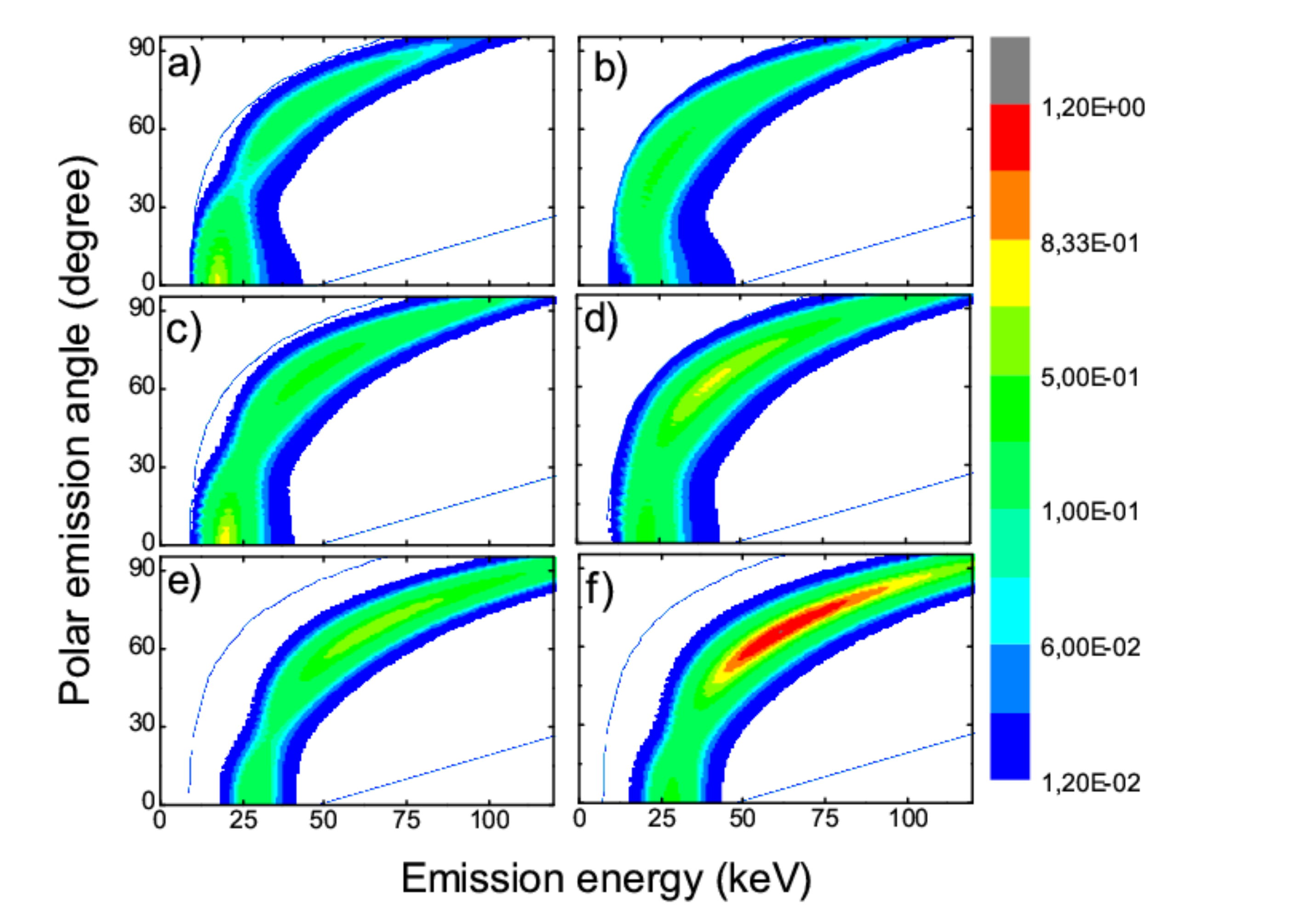}
\end{center}
\vspace{-0.5cm}  
\caption{ \footnotesize{ 
Same as in figure \ref{figure1}  
but the spectrum 
is given in the atomic (laboratory) frame 
in which the HCI moves under 
the polar angle of $180^{\circ}$. }}  
\label{figure3} 
\end{figure} 

\begin{figure}[t] 
\vspace{0.25cm}
\begin{center}
\includegraphics[width=0.5\textwidth]{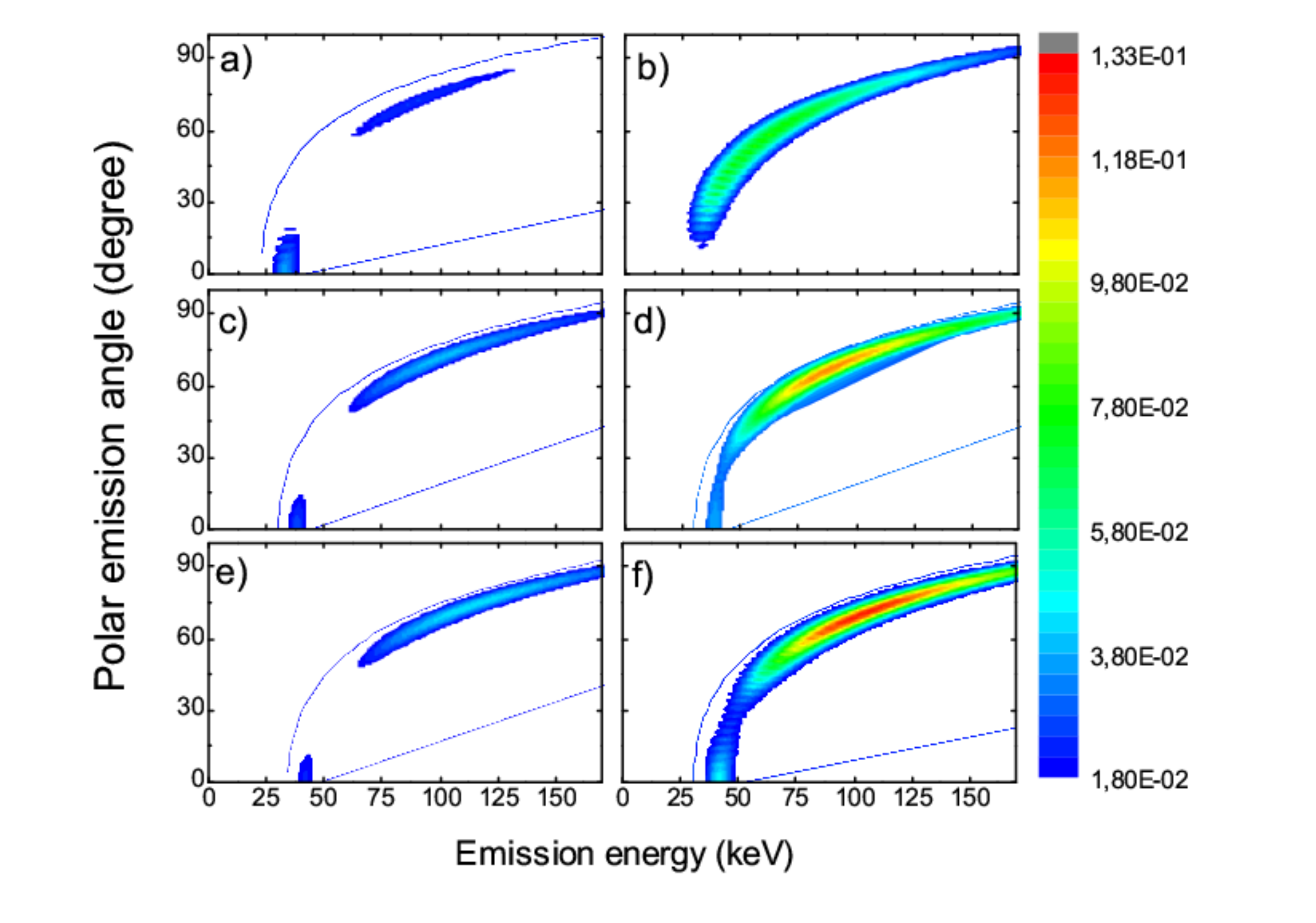}
\end{center}
\vspace{-0.5cm}  
\caption{ \footnotesize{ 
Same as in figure \ref{figure2}  
but the spectrum 
is given in the laboratory frame. }}  
\label{figure4} 
\end{figure} 

In figures \ref{figure3} and \ref{figure4} 
we present the emission spectrum 
for the correlated transfer ionization for the same collisions 
as in figures \ref{figure1} and \ref{figure2}, respectively.  
Now the spectrum is given in the laboratory frame in which 
the atom is at rest and the ion moves with a velocity 
${\bf v}_p =(0,0, - v)$ ($v >0$). 
It follows from the figures that in this frame the electron 
is emitted with a large energy in the direction opposite 
to the motion of the ion. The effect of the GBI is clearly 
seen also in the laboratory frame 
that makes it possible to verify it in experiment.  

In collisions with HCIs the so called independent transfer ionization yields the main contribution to the total transfer ionization. However, this channel is not of interest for the present study because the electron-electron interaction plays in it essentially no role since capture of one electron and emission of another proceed independently of each other. In the rest frame of the atom this channel leads to the emission of low energy electrons which preferably move in the direction of the motion of the ion. Therefore, in relativistic collisions the emission patterns due to the correlated and independent transfer ionization do not overlap.      

In conclusion, we have considered the correlated transfer ionization in relativistic collisions between a highly charged ion and a light atom. 
We have shown that this process is profoundly influenced 
by the generalized Breit interaction already at modest relativistic impact energies. This interaction qualitatively changes the shape of the emission pattern and strongly increases the emission. 
These effects can be verified experimentally by detecting  
high-energy electrons emitted in the laboratory frame 
under large angles ($ \sim 120^{\circ}$ - $180^{\circ}$) 
with respect to the motion of the highly charged ion.   

O.Y.A. and E.A.M. are grateful to the MPIK Heidelberg for the hospitality during their visits and to the DFG for financial support, project SU 658/2-1. The work of O.Y.A. and E.A.M. was supported by the RFBR grants 12-02-91340-NIIO$_a$, 14-02-00188A and by the Ministry of education and science of Russian Federation under the project 8420. 
The work of E.A.M. was also supported by German-Russian 
Interdisciplinary Science Center (G-RISC) grant P-2013a-1. 
A.B.V. acknowledges support 
from the Extreme Matter Institute EMMI and 
the DFG under the project VO 1278/2-1.


\begin{thebibliography}{99}
\bibitem{elect_pairs} 
A rough estimate for the contribution  
of the other pairs of electrons in a carbon atom 
can be obtained by considering effective localization 
volumes for these pairs (see \cite{2el-1ph-a}) 
which suggests that the other pairs yield about $25 \%$ 
contribution to the total cross section.
\bibitem{eet} The EECD should not be confused 
with the well known electron-electron Thomas mechanism of transfer ionization (L.H.Thomas, Proc.R.Soc. {\bf 114} 561 (1927)) which is effective in collisions with fast low charged ions ($Z_i \ll v \ll c$) 
and for which the interaction between the ion and only one of the atomic electrons is necessary. In nonrelativistic 
collisions with HCIs, when $Z_i \sim v$, the Thomas mechanism 
is already not possible kinematically. 
In the relativistic domain this mechanism 
is not supported by the kinematics 
both for $Z_i \ll v$ and $Z_i \sim v$.
\end{thebibliography}
\end{document}